\documentclass[a4paper, 11pt]{article}
\usepackage[affil-it]{authblk} 
\usepackage{etoolbox}
\usepackage{lmodern}
\usepackage[a4paper]{geometry}
\usepackage[in]{fullpage}
\usepackage[utf8]{inputenc}
\usepackage{amsmath}
\usepackage{amsfonts}
\usepackage{graphicx}
\usepackage[dvipsnames]{xcolor}
\usepackage[colorlinks]{hyperref}
\hypersetup{linkcolor=blue,citecolor=blue,urlcolor=blue}

\title{Unruh Duality and Maximum Acceleration in String Theory}

\makeatletter
\patchcmd{\@maketitle}{\LARGE \@title}{\fontsize{16}{19.2}\selectfont\@title}{}{}
\makeatother

\author[1,2]{Heliudson Bernardo\footnote{Email: \href{mailto:heliudson@hep.physics.mcgill.ca}{heliudson@hep.physics.mcgill.ca}}}

\affil[1]{Instituto de F\'isica Te\'orica, UNESP-Universidade Estadual Paulista, \protect\\
R. Dr. Bento T. Ferraz 271, Bl. II\\ S\~ao Paulo 01140-070, SP, Brazil}
\affil[2]{Department of Physics, McGill University,\protect\\ Montreal, QC, H3A 2T8, Canada}

\date{\vspace{-5ex}}

\begin{document}

\maketitle


\begin{abstract}
It is argued that the Unruh effect may be interpreted as a duality of a theory on different backgrounds. This issue is revisited in String Theory in the path integral formalism. By using T-duality and the Unruh effect, the T-dual transformation for acceleration is investigated and a maximum effective physical acceleration for observers in String Theory is found.

\end{abstract}



\section{Introduction}
\label{sec:intro}

String theory is a promising candidate for a physical theory of quantum gravity. It is based on the assumption that one-dimensional extended fundamental strings play a major role in the theory in many regimes. There is only one dimension-full parameter in the theory, the string length $\alpha'^{1/2}$, and all other parameters are vacuum expectation values of dynamical fields of the theory \cite{Polchinski:1998rq, Polchinski:1998rr}.

In some sense, this fundamental length is the minimum possible distance probed in String Theory. In scattering of strings at weak coupling, the hard limit of the amplitudes indicates objects of size $\alpha'^{1/2}$ \cite{Veneziano:1974dr}. Also, on toroidal compactifications of String Theory, there is an equivalence of large and small radii, when compared to the string length. This equivalence and its generalizations are referred to as T-duality between backgrounds (see \cite{Giveon:1994fu} and references therein). The self-dual radius, given exactly by the string length, yields a minimum effective physical length for the radii of the tori. 

Considering the low-energy limit of String Theory, a Supergravity solution is generally not supposed to hold in regions where the curvature is bigger than the string scale or at distances smaller than its characteristic length. In this case, corrections due to string-size effects are necessary. Usually, in any attempt to construct a field theory or to describe physical phenomena at energy scales closer to the inverse string length, new stringy effects appear \cite{Green:1987sp, Green:1987mn}.

Heuristically, relativistic theories with a minimum fundamental length $\lambda$ have an upper bound on the possible maximum acceleration $a_c$ given by
\begin{equation}
    a_c \sim \frac{1}{\lambda}.
\end{equation}
Extended objects have also a critical acceleration due to causality arguments. In fact, for an object with proper length $\lambda$, if some point on it has an acceleration bigger than $1/\lambda $ then it will be causally disconnected from other points on the object, since it will perceive a Rindler horizon at distance smaller than $\lambda$ \cite{Misner:1974qy}. So, consistent acceleration of fundamental extended objects should satisfy $a < 1/\lambda$. From this, one sees that accelerated fundamental strings should be treated carefully. One should also recall that in String Theory, the fundamental objects are $\textit{quantum}$ objects, such that simple classical heuristic arguments may not apply directly for fundamental strings. 

In fact, quantum mechanically, we also find possible limits on acceleration \cite{Caianiello1984}. For strings, in \cite{Sanchez:1989cw,Gasperini:1990pn} it was shown that arbitrary large acceleration leads to instabilities in the string proper's length. A universal maximum critical acceleration was first recognized as an intrinsic property of String Theory in \cite{Frolov:1990ct} (see also \cite{Gasperini:1992jz, Feoli:1992vmd,Martellini:1987ug,Sanchez:1991vv}). Also, strings in Rindler spacetime were considered in \cite{deVega:1987um}, motivated by the connection between accelerated frames and black hole backgrounds. A divergence in string partition functions was interpreted as maximal  acceleration close to Rindler horizon in \cite{Mertens:2015adr}, though the authors concluded that such divergence is due to an extrapolation of field theory to strings.

In any analysis of quantum relativistic systems on accelerated frames one should consider the Fulling-Davies-Unruh, Davies-Unruh, or simply Unruh effect \cite{Fulling:1972md,Davies:1974th, Unruh:1976db, Unruh:1983ac} (for a review, see \cite{Crispino:2007eb}). An observer accelerating in the inertial vacuum will be in a thermal bath with temperature that depends on its acceleration. So, quantization in its frame should be done taking this thermal bath into account. The Unruh effect in String Theory was considered in several papers from different perspectives \cite{deVega:1987um, Parentani:1989gq, Feoli:1993ew, Witten:2018xfj} (for connections with holography, see \cite{Paredes:2008cr}). A more systematic approach in terms of path integrals, analogous to the analysis of \cite{Unruh:1983ac} is presented in section \ref{sec3}.

Strings at finite temperature have interesting new properties when compared with particles \cite{Atick:1988si} (see also references therein). More striking is the existence of a maximum temperature for a gas of strings, the Hagedorn temperature\footnote{For investigations on string thermodynamics near the Hagedorn temperature and in Rindler spacetime, see \cite{Mertens:2013pza, Mertens:2013zya} and references therein.} $T_H$ \cite{Hagedorn:1965st}. Also, due to the existence of winding modes in the string case, T-duality implies that there is a duality between temperatures greater and smaller than a self-dual temperature of the order of $T_H$. Thus, effectively, there is a maximum possible temperature for a thermal bath of strings \cite{Atick:1988si}. From this observation and the connection between temperature and acceleration through the Unruh effect, we may suspect that there should possibly be a maximum acceleration for observers in String Theory. A relation between T-duality of temperature and acceleration seems very plausible at least.

The goal of this work is to study and elucidate the connection between the Unruh effect, acceleration, temperature and T-duality in String Theory. The paper is organized as follows: In section \ref{sec2} the Unruh effect as formulated in the path integral formalism is reviewed and it is argued that it may be seen as a duality (technical details are given in appendix \ref{appendix}); in section \ref{sec3}, a path integral proof of the Unruh effect first for relativistic point particle and then for strings is given; based on T-duality of temperature and the Unruh effect, a T-duality transformation of acceleration is proposed in section \ref{sec4}; section \ref{sec5} is a discussion of the results and a possible relation with the Hagedorn phase transition; in section \ref{sec6} the conclusions are presented.

\paragraph*{On notation.} Natural units are adopted throughout the paper, $c=1,\; \hbar=1,\; k_B =1$, and the subscript $E$ in some time variables is used to emphasize Euclidean signature.

\section{The Unruh effect as a duality}\label{sec2}

Quantum Field Theory is constructed in a Lorentz invariant way, such that the vacuum of a theory is the same for all inertial observers. The fact that two observers non-related by a Lorentz transformation assign different states for their respective vacua is the main reason for the Unruh Effect, which states that a uniform accelerated observer will experience a thermal bath in the vacuum  state of inertial observers. The temperature $T$ of the thermal state as seen by the accelerated observer is related with its proper acceleration $a$ by the Unruh temperature formula,
\begin{equation}\label{unruhtemperature}
    T= \frac{a}{2\pi}.
\end{equation}

In practice, one can show this by noticing that creation and annihilation operators in the mode expansion of fields in the accelerated frame are related to the corresponding operators in the inertial frame by a Bogoliubov transformation \cite{Crispino:2007eb}.

It was also shown in \cite{Unruh:1983ac} that the Unruh effect can be derived directly from the partition function of the theory. In fact, the Euclidean path integral for a field theory defined on a Rindler wedge, with Rindler metric
\begin{equation}
ds^2 = G_{\mu\nu}^{(a)}(x) dx^\mu dx^\nu= a^2r^2d\eta_E^2 + dr^2 +\delta_{ij}dx_{\perp}^idx_{\perp}^j,
\end{equation}
and with time compactified on a circle of radius $\beta/2\pi$ is identical to the path integral in Euclidean space provided $a\beta = 2\pi$, that is,
\begin{equation}\label{unruhduality}
    Z_{S^1_{\frac{\beta}{2\pi}}\times\mathbb{R}_{+}^{*} \times\mathbb{R}^{D-2}}\left[G_{\mu\nu} = G_{\mu\nu}^{(2\pi/\beta)}\right]= 
    Z_{\mathbb{R}^D}\left[G_{\mu\nu}= \delta_{\mu\nu}\right].
\end{equation}
The subscript in $Z$ is a reminder of where the coordinates appearing in the metric take values. For the Rindler wedge with Euclidean periodic time, we have $\left(\eta_E, r, x^i_\perp\right)\in S^1_{\beta/2\pi} \times\mathbb{R}_{+}^{*} \times\mathbb{R}$, as $r>0$, and the form of the Rindler metric implies that the theory in the left hand side of (\ref{unruhduality}) is defined on a cone (with deficit angle $2\pi - a \beta $) times flat directions. Thus, for $a\beta = 2\pi$ it is topologically equivalent to a flat space.

Since the Euclidean path integral with time taking values in $S^1_{\beta/2\pi}$ defines a field theory at finite temperature with temperature $T=1/\beta$, the condition on the acceleration for the Rindler metric in the left side of equation (\ref{unruhduality}) is equivalent to equation (\ref{unruhtemperature}). Therefore, equation (\ref{unruhduality}) is a formal way to state the Unruh effect: the Euclidean generator for correlators for the accelerated observer is the same as the inertial one provided the thermal bath is taken into account \emph{and} its temperature is related with the proper acceleration via (\ref{unruhtemperature}).

These results were proved for any interacting field theory containing scalars and spin-$1/2$ fields, and in the presence of sources in \cite{Unruh:1983ac}. For a review of the calculation, see appendix \ref{appendix}. So, \textit{if} $a \beta =2\pi$, any correlation function to be measured by the accelerated observer at finite temperature should be equal to the correlation functions that inertial observers would measure\footnote{The local operators in the correlators of both theories should be evaluated at the corresponding spacetime points for both observers. Also, one should consider the relative transformation of the operators for each observer.}.

Generally, the equality of partition functions implies that the theories are the same or that at least they are dual to each other. Having theories in different backgrounds giving the same partition function is the definition of a duality between the backgrounds. So, from equation (\ref{unruhduality}) we conclude that the Unruh effect may be seen as a duality between different backgrounds for the same relativistic quantum theory.

\section{Finite temperature String Theory on the Rindler wedge}\label{sec3}

Similar to what was done for scalar and spin-$1/2$ fields in \cite{Unruh:1983ac} (see also appendix \ref{appendix}), we can consider the string partition function on a Rindler wedge with compactified time and prove the Unruh effect in String Theory in the path integral formalism. Before doing that, let us consider the relativistic point particle case,
\begin{align}
    Z_{S^1_{\frac{\beta}{2\pi}}\times\mathbb{R}_{+}^{*} \times\mathbb{R}^{D-2}}\left[G_{\mu\nu}^{(a)}\right] =  \int [dXde]_R &\exp\left\{-\frac{1}{2}\int d\tau \left[e^{-1}\left(a^2r^2{\Dot{\eta}_E}^2 +\right.\right.\right.\nonumber\\
    &+\left.\left.\left.\Dot{r}^2+ \delta_{ij}\Dot{X}^i_{\perp}\Dot{X}^j_{\perp}\right)+em^2\right]\vphantom{\frac{1}{2}}\right\},
\end{align}
where $e(\tau)$ is the einbein of the worldline $X^\mu(\tau)=(\eta_E(\tau), r(\tau), X^i_{\perp}(\tau))$ parameterized by $\tau$ and overdots denotes derivation with respect to this parameter. The subscript $R$ indicates that the path integral is over worldlines in the compactfied Rindler wedge, i.e., $\left(\eta_E(\tau), r(\tau)\right) \in S^1_{\beta/2\pi}\times \mathbb{R}_{+}^{*}$. 

Making the following change of variables in the path integral,
\begin{equation}
    T(\tau) = r(\tau)\sin(a\eta_E(\tau)), \quad X(\tau) = r(\tau)\cos(a\eta(\tau)),
\end{equation}
gives
\begin{align}
    Z_{S^1_{\frac{\beta}{2\pi}}\times\mathbb{R}_{+}^{*} \times\mathbb{R}^{D-2}}\left[G_{\mu\nu}^{(a)}\right] = \int [dXde]_R&\exp\left\{-\frac{1}{2}\int d\tau \left[e^{-1}\left(\Dot{T}^2+ \Dot{X}^2 + \delta_{ij}\Dot{X}^i_{\perp}\Dot{X}^j_{\perp}\right) em^2 \vphantom{\frac{1}{2}}\right]\right\} .    
\end{align}
Due to the periodicity of $\eta_E$, not all the $(T,X)$ plane is accessible for the integrated worldlines. But for the particular value $\beta = 2\pi/a$, the path integral in the new variables is over worldlines in all coordinate space, that is, for this specific value of $\beta$ we have $-\infty <T,X,X^i_{\perp}< \infty$. In this case we can drop the subscript $R$ and write
\begin{align}
    Z_{S^1_{\frac{\beta}{2\pi}}\times\mathbb{R}_{+}^{*} \times\mathbb{R}^{D-2}}\left[G_{\mu\nu}^{(a= 2\pi/\beta)}\right] &= \int [dXde]\exp\left\{-\frac{1}{2}\int d\tau \left[e^{-1}\delta_{\mu\nu}\Dot{X}^\mu\Dot{X}^\nu+em^2\right]\right\}\nonumber\\
    &= Z_{\mathbb{R}^D}\left[\delta_{\mu\nu}\right],
\end{align}
which proves the Unruh duality for a relativistic point particle.

Similarly, we can consider the string partition function on a Rindler wedge with compactfied time,
\begin{equation}
    Z_{S^1_{\frac{\beta}{2\pi}}\times\mathbb{R}_{+}^{*} \times\mathbb{R}^{D-2}}\left[G_{\mu\nu}^{(a)}\right] = \int [dX d\gamma]_R \exp\left\{-\frac{1}{4\pi\alpha'}\int d^2\sigma \sqrt{-\gamma}\gamma^{ab}\partial_a X^\mu \partial_b X^\nu G_{\mu\nu}^{(a)} \right\},
\end{equation}
where the subscript $R$ in the functional measure is a reminder that we are summing over embeddings of the worldsheet onto the compactified Rindler wedge, i.e., that the target space function's components take values in $S^1_{\beta/2\pi}\times\mathbb{R}_{+}^{*} \times\mathbb{R}^{D-2}$. We have
\begin{align}
    Z_{S^1_{\frac{\beta}{2\pi}}\times\mathbb{R}_{+}^{*} \times\mathbb{R}^{D-2}}\left[G_{\mu\nu}^{(a)}\right] = \int [dX d\gamma]_R \exp&\left\{-\frac{1}{4\pi\alpha'}\int d^2\sigma  \sqrt{-\gamma}\gamma^{ab}\left[a^2r^2\partial_a \eta_E \partial_b \eta_E + \right.\right.\nonumber\\ &+\left.\left.\partial_a r\partial_b r + \delta_{ij}\partial_a X_{\perp}^i \partial_b X^j_{\perp} \right]\vphantom{\frac{1}{2}} \right\},
\end{align}
with
\begin{equation}
    \quad 0< \eta< \beta, \quad 0 < r< \infty, \quad -\infty< X^i_{\perp}<\infty.
\end{equation}

Making the following change of variables
\begin{align}
    X(\sigma) &= r(\sigma)\cos{(a\eta(\sigma))},\\
    T(\sigma) &= r(\sigma)\sin{(a\eta(\sigma))},
\end{align}
we get
\begin{align}
    Z_{S^1_{\frac{\beta}{2\pi}}\times\mathbb{R}_{+}^{*} \times\mathbb{R}^{D-2}}\left[G_{\mu\nu}^{(a)}\right] = \int [dX d\gamma]_R \exp&\left\{-\frac{1}{4\pi\alpha'}\int d^2\sigma  \sqrt{-\gamma}\gamma^{ab}\left[\partial_a T \partial_b T +\right.\right.\\
    &+\left.\left.\partial_a X \partial_b X + \delta_{ij}\partial_a X_{\perp}^i \partial_b X^j_{\perp} \right] \right\}\nonumber\\
    =  \int [dX d\gamma]_R \exp&\left\{-\frac{1}{4\pi\alpha'}\int d^2\sigma \sqrt{-\gamma}\gamma^{ab}\partial_a X^\mu \partial_b X^\nu \delta_{\mu\nu} \right\}.
\end{align}
If we now consider $a\beta =2\pi$, as in the point particle case, the new variables cover the entire target space and we have\footnote{This result was also explored in \cite{Mertens:2016tqv} by explicitly computing the partition functions.}
\begin{equation}\label{thermalstringduality}
    Z_{S^1_{\frac{\beta}{2\pi}}\times\mathbb{R}_{+}^{*} \times\mathbb{R}^{D-2}}\left[G_{\mu\nu}^{(a = \frac{2\pi}{\beta})}\right] =Z_{\mathbb{R}^D}\left[\delta_{\mu\nu}\right].
\end{equation}

So, the Polyakov path integral for an accelerating observer in a stringy thermal bath is equal to the Polyakov path integral for an inertial observer \emph{if} the acceleration and temperature are related by the Unruh temperature formula (\ref{unruhtemperature}). Notice that this case is physically different of having an accelerating string as seen by an inertial observer.

\section{T-duality transformation for acceleration}\label{sec4}

Let us step back and consider strings at finite temperature as seen by inertial observers. This case is different from having finite temperature strings for accelerated observers with a Rindler metric, that was considered in the previous section; as the vacuum state for inertial observers appears as a thermal state for the accelerated observers, a thermal state for inertial observers \textit{is not} a thermal state for the accelerated ones.

For the finite temperature strings in Minskowski spacetime, since the thermal partition function has a compactified Euclidean direction, we can do a T-duality transformation on this direction and this should leave the string partition function invariant. Indeed, in the inertial background, this may be used to find the T-duality transformation of temperature, the radius $R = \beta/2\pi$ transforming as
\begin{equation}
    \frac{\beta}{2\pi} \rightarrow \frac{\Tilde{\beta}}{2\pi} = \frac{2\pi}{\beta}\alpha' \implies \Tilde{\beta} = \frac{4\pi^2\alpha'}{\beta}.    
\end{equation}
This implies that the T-dual temperature is (see chapter 9 in \cite{Polchinski:1998rq})
\begin{equation}\label{tdualitytemperature}
    \Tilde{T} = \frac{4 T^2_H}{T},
\end{equation}
where $T_H \equiv 1/(4\pi\alpha'^{1/2})$ is the Hagedorn temperature \cite{Atick:1988si}. From this T-duality transformation we can calculate the self-dual temperature $T_*$, given by
\begin{equation}
    T_* = 2T_H = \frac{1}{2\pi \alpha'^{1/2}}.
\end{equation}

Now, consider two accelerated observers with different constant proper accelerations, but such that the corresponding thermal baths have T-dual temperatures. If String Theory describe their physics, the observers would never differ between these two temperatures, as both are physically equivalent. In a sense, they would be ``dual" to each other, as their measurements of physical quantities associated with the thermal baths would be the same. Thus, by the relation between temperature and acceleration, it is possible to define a ``dual" acceleration.

To find how an observer's acceleration ``transforms" under T-duality, we use the Unruh duality, equation (\ref{thermalstringduality}) twice: for the observer that measures a thermal bath at temperature $T$ and then for the observer at dual temperature $\Tilde{T}$, 
\begin{align}\label{unruhstring}
    Z_{S^1_{\frac{\beta}{2\pi}}\times\mathbb{R}_{+}^{*} \times\mathbb{R}^{D-2}}\left[G_{\mu\nu}^{(a= 2\pi/\beta)}\right] &= Z_{\mathbb{R}^D}\left[\eta_{\mu\nu}\right]= Z_{S^1_{\frac{\Tilde{\beta}}{2\pi}}\times\mathbb{R}_{+}^{*} \times\mathbb{R}^{D-2}}\left[G_{\mu\nu}^{(\Tilde{a})}\right],
\end{align}
so we should have
\begin{equation}
    \Tilde{a} = \frac{2\pi}{\Tilde{\beta}} = \frac{\beta}{2\pi\alpha'} = \frac{1}{a\alpha'}.
\end{equation}
Therefore, we found the T-duality transformation for observer's acceleration,
\begin{equation}
    a \rightarrow \Tilde{a} = \frac{1}{a\alpha'}.
\end{equation}
The self-dual acceleration is given by
\begin{equation}
    a_*= \frac{1}{\alpha'^{1/2}},
\end{equation}
and it corresponds to the self-dual temperature by (\ref{unruhtemperature}).

So, from Unruh duality and T-duality, we conclude that the thermal state for the observer with acceleration $a$ is a thermal state with T-dual temperature for the observer with acceleration $\Tilde{a}$. Since their respective partition functions are equal, both observers would measure the same correlators and so amplitudes, upon considering their respective thermal baths with temperatures related by T-duality. As far as string amplitudes are concerned,
they could not distinguish between being accelerated at uniform acceleration $a$ and in thermal bath at temperature $T$ or being accelerated at uniform acceleration $1/(a\alpha')$ and in thermal bath at temperature $4T^2_H/T$, by probing \textit{only} their thermal baths. This is in the same level as physical equivalence of compactifications on tori with radius $R$ or $\alpha'/R$.

Therefore, we conclude that effectively, there is a maximal possible acceleration in String Theory, given by the self-dual value $a_*$, on the same grounds as there is an effective minimal length given by the self-dual radius $R_*= \alpha'^{1/2}$.

\section{Discussion}\label{sec5}

T-duality is a duality between different but equivalent backgrounds in String Theory. The results in this paper indicates that T-duality may also be used to relate measurements of different observers in String Theory, a property that seems not to have been explored before. Notice that this "application" of T-duality depends more on the state of motion of the T-dual-related observers than on the geometry of the background. In the case presented here, we were able to find the "T-dual motion" of proper acceleration. Generalizations of this case are under investigation.

It is worth emphasizing that the equalities in equation (\ref{unruhstring}) are \textit{not} due to T-duality transformation of the metric. In fact, using Buscher's rules \cite{Buscher:1987sk,Buscher:1987qj} to find out the T-dual Rindler metric in the $S^1_{\beta/2\pi}$ direction, we would get
\begin{equation}
    d\Tilde{s}^2 = \frac{1}{a^2r^2}d\Tilde{\eta}_E^2 + dr^2 +\delta_{ij}dx_{\perp}^idx_{\perp}^j, 
\end{equation}
with $\Tilde{\eta}_E \in S^1_{\Tilde{\beta}/2\pi}$, which is not a Rindler metric for acceleration $1/a$. But the proper times of the observers with accelerations connected by T-duality are the same, and this indicates that they are related\footnote{Also, even considering $a\beta\neq2\pi$, it is not clear if the fact that the Rindler coordinates do not cover the entire space plays a big role in ``T-dualizing'' observers.}. Regardless of such observations, T-duality was only used to construct the thermal bath with T-dual temperature and to argue that the physics in the presence of both thermal baths is the same.

Since thermal baths in String Theory were considered, one possible question is the relation of the present work with the Hagedorn phase transition \cite{Atick:1988si}. It is known that the thermal partition function for a string gas diverges at temperatures equal or greater than Hagedorn temperature, physically due to the exponential energy growth in the density of string states. This divergence is believed to signal a first order phase transition, not totally understood (see also \cite{Antoniadis:1991kh}). But the temperature due to the Unruh effect cannot give rise to phase transitions, since the thermal state as seen by the accelerated observer is just the vacuum state of the inertial observers. Accelerated fields or strings were not considered in \cite{Unruh:1983ac} or in the current work. What was done in this paper is different than calculating the thermal partition function for a gas of strings in Minskowski spacetime. Only the usual Polyakov path integral for an inertial string was used, but as seen by different observers. So, this work does not offer any new approach for the Hagedorn phase transition.

A potential problem in the arguments of the previous section is the fact that the self-dual temperature $T_*$ is larger than the Hagedorn temperature and so its usage is subtle. But note that the value of the self-dual temperature comes from the T-duality transformation formula, equation (\ref{tdualitytemperature}), that holds as long as the target space has an compactified Euclidean time direction. Also, there are thermal string constructions free of Hagedorn instabilities (see for example \cite{Florakis:2010is,Kounnas:2011fk, PandoZayas:2002hh, Chaudhuri:2014hoa}) and the arguments of the previous section can be applied for the bosonic sector of them. Also, there could be potential applications to Euclidean type $\text{II}^*$ theories\footnote{The author is grateful to Keshav Dasgupta for bringing that to his attention.} (see \cite{Hull:1998vg} and references therein), as they can be seen as T-dual to type II strings (though in the time-like direction). 

Notice that the arguments to find the maximum acceleration presented in the previous section are generally valid, due to the universality of T-duality and the Unruh effect. In particular, the relation between  the maximum acceleration $a_{\text{max}}$ and the self-dual temperature,
\begin{equation}
    a_{\text{max}} = 2\pi T_*,
\end{equation}
is expected to hold true in general and it is the main result of the present work.

\section{Conclusion}\label{sec6}

In this paper, a novel proof of the Unruh effect in String Theory was given, in the path integral formalism. Using this result and T-duality, a T-dual transformation of an observer's proper acceleration was found and it was argued that there is an effective maximum acceleration in String Theory, as would be heuristically expected from a theory with minimal length. All of these results were obtained in bosonic String Theory. Generalizations of this particular case are under investigation, even though only mild modifications are physically expected.

\section*{Acknowledgments}
The author would like to thank J\'essica Martins, George Matsas, Gabriel Cozzella, Yigit Yargic and Guilherme Franzmann for interesting discussions and Robert Brandenberger, Horatiu Nastase and Keshav Dasgupta for useful comments on previous versions of the manuscript. This work is fully supported by CAPES-Brazil. The author is also thankful to McGill University for hospitality during the execution of this work.

\appendix
\section{Unruh duality in field theory}\label{appendix}

In \cite{Unruh:1983ac} it was shown that all vacuum Green's functions between spacetime points within the same Rindler wedge for inertial observers are the same as the Green's functions for accelerated observers in thermal equilibrium at a temperature $T = a/2\pi$. This was done starting from the phase space path integral for scalar and spin-$1/2$ fields. In this appendix, the core of such results is reviewed, by showing equation (\ref{thermalstringduality}) for spin-0 and spin-$1/2$ theories.

Let us consider scalar fields first. The starting point is the Euclidean partition function for the accelerated observers:
\begin{equation}
    Z^R(\beta) = \int [d\phi]_P\exp\left\{-S_E^R(\beta)\right\},
\end{equation}
where
\begin{align}
    S^R_E(\beta) = \int_0^{\beta}d\eta_E\int_{r>0}dr d^{D-2}x_{\perp}\sqrt{G^E}\left[\frac{1}{2}(G^E)^{\alpha \beta}\partial_{\alpha}\phi\partial_{\beta}\phi + V(\phi)\right],
\end{align}
with
\begin{equation}
(ds^E)^2 = G^E_{\alpha \beta}dx^{\alpha}dx^{\beta}= a^2r^2d\eta_E^2 + dr^2+ \delta_{ij}dx_{\perp}^idx_{\perp}^j. 
\end{equation}
The superscript $R$ indicates that the region of integration in the action is over the Rindler wedge and the subscript $P$ denotes that the path integral is over field configurations with periodic boundary condition in time with period $\beta$, ${\phi(\eta_E=0)=\phi(\eta_E=\beta)}$. Changing variables to 
\begin{equation}\label{changeofvar}
    t_E = r\sin(a\eta_E), \qquad x = r\cos(a\eta_E),
\end{equation}
with $\beta a \leq 2\pi$ (in order to this transformation be single valued), we have
\begin{equation}
    (ds^E)^2 = \delta_{\mu\nu}dx^{\mu}dx^{\nu}=dt_E^2 + dx^2 + \delta_{ij}dx_{\perp}^idx_{\perp}^j,
\end{equation}
and the Euclidean path integral is 
\begin{align}
    Z^R(\beta) = \int[d\phi]_P\exp\left\{- \int_R dt_Edx d^{D-2}x_{\perp}\left[\frac{1}{2}\delta^{\mu\nu}\partial_{\mu}\phi \partial_{\nu}\phi + V(\phi)\right]\right\}. 
\end{align}

The periodic boundary condition becomes
\begin{equation}
    \phi(0) = \phi(r\sin(a\beta)),
\end{equation}
and then, if $\beta = 2\pi/a$, it turns into a consistency condition. Also, for this value of $\beta$, the region $R$ is the full $(t_E,x,x_{\perp}^i)$ space and so we get
\begin{equation}
    Z^R(\beta = 2\pi/a) = \int [d\phi] \exp \left\{-S_E\right\},
\end{equation}
which is the Euclidean generating functional for the theory in inertial coordinates, i.e., for inertial observers. This proves Unruh duality for scalar theories.

For spin-$1/2$ case, we need to consider the Dirac action in Rindler coordinates. Let us write the action for a Dirac fermion in covariant form,
\begin{equation}
    S = -\int d^Dx e(x)\Bar{\psi}(x)(\gamma^a\nabla_a+m)\psi(x),
\end{equation}
where $e = \det e_\mu^{\;\;a}$ is the determinant of the vielbein fields $e_{\mu}^{\;\;a}$, $\gamma^a$ are the Dirac matrices and
\begin{equation}
    \gamma^a\nabla_a = \gamma^ce_c^{\;\;\mu}\left(\partial_\mu - \frac{i}{2}(\omega_\mu)^{ab}\Sigma_{ab}\right),
\end{equation}
with $(\omega_\mu)^{ab}$ the spin connection and $\Sigma^{ab} = -i/4[\gamma^a,\gamma^b]$ the Lorentz generators in spin-$1/2$ representation. For the Rindler metric, the only non-vanishing independent component of the spin connection is
\begin{equation}
    \omega^0_{\;\;1} = ad\eta,
\end{equation}
such that the Dirac action on the Rindler wedge $r> 0$ is
\begin{align}
    S^R= -\int_{r>0}d\eta dr d^{D-2}x_{\perp}ar\Bar{\psi}\left(\frac{1}{ar}\gamma^0 \partial_{\eta} + \gamma^1\partial_r  + \gamma^i\partial_i+\frac{1}{2r}\gamma^1 + m\right)\psi.
\end{align}

Now, consider the thermal partition function for the accelerated observers,
\begin{equation}
    Z^R(\beta) = \int[d\Bar{\psi}d\psi]_{\psi(0) = -\psi(\eta_E = \beta)}\exp\left\{-S_E^R(\beta)\right\},
\end{equation}
where the fermion fields have antiperiodic boundary condition in time with period $\beta$ and
\begin{align}
    S_E^R(\beta) = \int^{\beta}_0d\eta_E\int_{r>0}dr d^{D-2}x_{\perp}ar \Bar{\psi}\left(\frac{i}{ar}\gamma^0 \partial_{\eta_E} +\gamma^1\partial_r  + \gamma^i\partial_i+\frac{1}{2r}\gamma^1 + m\right)\psi,
\end{align}
is the Euclidean Dirac action in Rindler coordinates. Performing a change of variables in the functional path integration by
\begin{align}
    \psi(x) \rightarrow  M(\eta_E)\psi(x) &\equiv \left(\cos(a\eta_E/2)+ i\gamma^0\gamma^1\sin(a\eta_E/2)\right)\psi(x),\\
    \Bar{\psi}(x) \rightarrow \Bar{\psi}(x)M^{\dagger}(\eta_E) &\equiv \Bar{\psi}(x)\left(\cos(a\eta_E/2)- i\gamma^0\gamma^1\sin(a\eta_E/2)\right),
\end{align}
leads to 
\begin{align}
    Z^R(\beta) = \int[d\Bar{\psi}d\psi]_A &\exp\left\{-\int_0^\beta d\eta_E \int_{r>0} dr d^{D-2}x_{\perp} ar \left[\frac{i}{ar}\Bar{\psi}\Bar{\Sigma}^0 \partial_{\eta_E}\psi + \Bar{\psi}\Bar{\Sigma}^1\partial_r\psi  +\right.\right.\nonumber\\
    &+\left.\left.\Bar{\psi}\gamma^i\partial_i\psi+ m\Bar{\psi}\psi\right]\vphantom{\frac{1}{2}}\right\}.
\end{align}
The subscript $A$ in the functional measure indicates the antiperiodic condition that is now written as
\begin{equation}
    \psi(\eta_E= 0) = - \left(\cos(a\beta/2)+ i\gamma^0\gamma^1\sin(a\beta/2)\right)\psi(\eta_E = \beta).
\end{equation}

Using the same change of variable as in the scalar case equation (\ref{changeofvar}), after some algebra we get
\begin{equation}
    Z^R(\beta) = \int[d\Bar{\psi}d\psi]_A\exp\left\{-\int_Rdt_Edxd^{D-2}x_{\perp}\Bar{\psi}\left[i\gamma^0\partial_{t_E}+ \gamma^1\partial_x + \gamma^i \partial_i +m\right]\psi \right\}.
\end{equation}
If now we take $\beta =2\pi/a$, the antiperiodic condition becomes a consistency condition and the region $R$ turns into all space in $(t_E,x,x^i_{\perp})$ coordinates. Thus,
\begin{equation}
    Z^R(\beta = 2\pi/a) = \int[d\Bar{\psi}d\psi]\exp\left\{-S_E \right\},
\end{equation}
where $S_E$ is the Dirac action in inertial coordinates. This is the Unruh duality for spin-$1/2$ fields.

In \cite{Unruh:1983ac} it is argued that we can use the results presented in this appendix to prove the Unruh effect for any interacting field theory involving scalars fields and Dirac spinors.



\bibliographystyle{unsrt} 
\bibliography{References}

\begin{thebibliography}{10}

\bibitem{Polchinski:1998rq}
J.~Polchinski.
\newblock {\em {String theory. Vol. 1: An introduction to the bosonic string}}.
\newblock Cambridge Monographs on Mathematical Physics. Cambridge University
  Press, 2007.

\bibitem{Polchinski:1998rr}
J.~Polchinski.
\newblock {\em {String theory. Vol. 2: Superstring theory and beyond}}.
\newblock Cambridge Monographs on Mathematical Physics. Cambridge University
  Press, 2007.

\bibitem{Veneziano:1974dr}
G.~Veneziano.
\newblock {An Introduction to Dual Models of Strong Interactions and Their
  Physical Motivations}.
\newblock {\em Phys. Rept.}, 9:199--242, 1974.

\bibitem{Giveon:1994fu}
Amit Giveon, Massimo Porrati, and Eliezer Rabinovici.
\newblock {Target space duality in string theory}.
\newblock {\em Phys. Rept.}, 244:77--202, 1994.

\bibitem{Green:1987sp}
Michael~B. Green, J.~H. Schwarz, and Edward Witten.
\newblock {\em {Superstring Theory. Vol. 1: Introduction}}.
\newblock Cambridge Monographs on Mathematical Physics. 1988.

\bibitem{Green:1987mn}
Michael~B. Green, J.~H. Schwarz, and Edward Witten.
\newblock {\em {Superstring Theory. Vol. 2: Loop Amplitudes, Anomalies and
  Phenomenology}}.
\newblock Cambridge Monographs on Mathematical Physics. 1988.

\bibitem{Misner:1974qy}
Charles~W. Misner, K.~S. Thorne, and J.~A. Wheeler.
\newblock {\em {Gravitation}}.
\newblock W. H. Freeman, San Francisco, 1973.

\bibitem{Caianiello1984}
E.~E. Caianiello.
\newblock Maximal acceleration as a consequence of heisenberg's uncertainty
  relations.
\newblock {\em Lettere al Nuovo Cimento (1971-1985)}, 41(11):370--372, Nov
  1984.

\bibitem{Sanchez:1989cw}
Norma~G. Sanchez and G.~Veneziano.
\newblock {Jeans Like Instabilities for Strings in Cosmological Backgrounds}.
\newblock {\em Nucl. Phys.}, B333:253--266, 1990.

\bibitem{Gasperini:1990pn}
M.~Gasperini.
\newblock {Kinematic interpretation of string instability in a background
  gravitational field}.
\newblock {\em Phys. Lett.}, B258:70--74, 1991.

\bibitem{Frolov:1990ct}
Valeri~P. Frolov and Norma~G. Sanchez.
\newblock {Instability of Accelerated Strings and the Problem of Limiting
  Acceleration}.
\newblock {\em Nucl. Phys.}, B349:815--838, 1991.

\bibitem{Gasperini:1992jz}
M.~Gasperini.
\newblock {Causal horizons, accelerations and strings}.
\newblock {\em Gen. Rel. Grav.}, 24:219--223, 1992.

\bibitem{Feoli:1992vmd}
A.~Feoli and G.~Scapetta.
\newblock {Accelerated Strings with Limited Proper Acceleration}.
\newblock In {\em {Proceedings, International Conference on Structure: from
  Physics to General Systems}}, pages 107--117, 1992.

\bibitem{Martellini:1987ug}
M.~Martellini and Norma~G. Sanchez.
\newblock {Thermal event horizons, moduli and boson operator formalism on
  Riemann surfaces}.
\newblock {\em Phys. Lett.}, B192:361--367, 1987.

\bibitem{Sanchez:1991vv}
Norma~G. Sanchez.
\newblock {Conceptual aspects of string theory and the problem of maximal
  acceleration}.
\newblock In {\em {Proceedings, International Conference on Structure: from
  Physics to General Systems}}, pages 118--137, 1992.

\bibitem{deVega:1987um}
H.~J. de~Vega and Norma~G. Sanchez.
\newblock {String Quantization in Accelerated Frames and Black Holes}.
\newblock {\em Nucl. Phys.}, B299:818, 1988.

\bibitem{Mertens:2015adr}
Thomas~G. Mertens, Henri Verschelde, and Valentin~I. Zakharov.
\newblock {Revisiting noninteracting string partition functions in Rindler
  space}.
\newblock {\em Phys. Rev.}, D93(10):104028, 2016.

\bibitem{Fulling:1972md}
Stephen~A. Fulling.
\newblock {Nonuniqueness of canonical field quantization in Riemannian
  space-time}.
\newblock {\em Phys. Rev.}, D7:2850--2862, 1973.

\bibitem{Davies:1974th}
P.~C.~W. Davies.
\newblock {Scalar particle production in Schwarzschild and Rindler metrics}.
\newblock {\em J. Phys.}, A8:609--616, 1975.

\bibitem{Unruh:1976db}
W.~G. Unruh.
\newblock {Notes on black hole evaporation}.
\newblock {\em Phys. Rev.}, D14:870, 1976.

\bibitem{Unruh:1983ac}
William~G. Unruh and Nathan Weiss.
\newblock {Acceleration Radiation in Interacting Field Theories}.
\newblock {\em Phys. Rev.}, D29:1656, 1984.

\bibitem{Crispino:2007eb}
Luis C.~B. Crispino, Atsushi Higuchi, and George E.~A. Matsas.
\newblock {The Unruh effect and its applications}.
\newblock {\em Rev. Mod. Phys.}, 80:787--838, 2008.

\bibitem{Parentani:1989gq}
Renaud Parentani and Robertus Potting.
\newblock {The Accelerating Observer and the Hagedorn Temperature}.
\newblock {\em Phys. Rev. Lett.}, 63:945, 1989.

\bibitem{Feoli:1993ew}
A.~Feoli.
\newblock {String dynamics in Rindler space in a model with maximal
  acceleration}.
\newblock {\em Nucl. Phys.}, B396:261--269, 1993.

\bibitem{Witten:2018xfj}
Edward Witten.
\newblock {Open Strings On The Rindler Horizon}.
\newblock {\em JHEP}, 01:126, 2019.

\bibitem{Paredes:2008cr}
Angel Paredes, Kasper Peeters, and Marija Zamaklar.
\newblock {Temperature versus acceleration: The Unruh effect for holographic
  models}.
\newblock {\em JHEP}, 04:015, 2009.

\bibitem{Atick:1988si}
Joseph~J. Atick and Edward Witten.
\newblock {The Hagedorn Transition and the Number of Degrees of Freedom of
  String Theory}.
\newblock {\em Nucl. Phys.}, B310:291--334, 1988.

\bibitem{Mertens:2013pza}
Thomas~G. Mertens, Henri Verschelde, and Valentin~I. Zakharov.
\newblock {Near-Hagedorn Thermodynamics and Random Walks: a General Formalism
  in Curved Backgrounds}.
\newblock {\em JHEP}, 02:127, 2014.

\bibitem{Mertens:2013zya}
Thomas~G. Mertens, Henri Verschelde, and Valentin~I. Zakharov.
\newblock {Random Walks in Rindler Spacetime and String Theory at the Tip of
  the Cigar}.
\newblock {\em JHEP}, 03:086, 2014.

\bibitem{Hagedorn:1965st}
R.~Hagedorn.
\newblock {Statistical thermodynamics of strong interactions at high-energies}.
\newblock {\em Nuovo Cim. Suppl.}, 3:147--186, 1965.

\bibitem{Mertens:2016tqv}
Thomas~G. Mertens, Henri Verschelde, and Valentin~I. Zakharov.
\newblock {String Theory in Polar Coordinates and the Vanishing of the One-Loop
  Rindler Entropy}.
\newblock {\em JHEP}, 08:113, 2016.

\bibitem{Buscher:1987sk}
T.~H. Buscher.
\newblock {A Symmetry of the String Background Field Equations}.
\newblock {\em Phys. Lett.}, B194:59--62, 1987.

\bibitem{Buscher:1987qj}
T.~H. Buscher.
\newblock {Path Integral Derivation of Quantum Duality in Nonlinear Sigma
  Models}.
\newblock {\em Phys. Lett.}, B201:466--472, 1988.

\bibitem{Antoniadis:1991kh}
Ignatios Antoniadis and C.~Kounnas.
\newblock {Superstring phase transition at high temperature}.
\newblock {\em Phys. Lett.}, B261:369--378, 1991.

\bibitem{Florakis:2010is}
Ioannis Florakis, Costas Kounnas, Herve Partouche, and Nicolaos Toumbas.
\newblock {Non-singular string cosmology in a 2d Hybrid model}.
\newblock {\em Nucl. Phys.}, B844:89--114, 2011.

\bibitem{Kounnas:2011fk}
Costas Kounnas, Herve Partouche, and Nicolaos Toumbas.
\newblock {Thermal duality and non-singular cosmology in d-dimensional
  superstrings}.
\newblock {\em Nucl. Phys.}, B855:280--307, 2012.

\bibitem{PandoZayas:2002hh}
Leopoldo~A. Pando~Zayas and Diana Vaman.
\newblock {Strings in RR plane wave background at finite temperature}.
\newblock {\em Phys. Rev.}, D67:106006, 2003.

\bibitem{Chaudhuri:2014hoa}
Shyamoli Chaudhuri.
\newblock {Euclidean time formulation for the superstring ensembles:
  Perturbative canonical ensemble with Neveu-Schwarz $B$-field backgrounds}.
\newblock {\em Phys. Rev.}, D90(12):126005, 2014.

\bibitem{Hull:1998vg}
C.~M. Hull.
\newblock {Timelike T duality, de Sitter space, large N gauge theories and
  topological field theory}.
\newblock {\em JHEP}, 07:021, 1998.

\end{thebibliography}





\end{document}